\documentclass[pre,twocolumn,superscriptaddress,showpacs,showkeys]{revtex4-1}
\usepackage{float}
\usepackage{times}
\usepackage[dvips]{color}
\usepackage{mathtools}
\usepackage{dcolumn}
\usepackage{bm}
\usepackage{amsmath}
\usepackage{amssymb}
\usepackage{hyperref}

\renewcommand{\d}{\mathrm{d}}
\newcommand{\ds}{D_{\mathrm{s}}}
\newcommand{\dm}{D_{\mathrm{m}}}
\newcommand{\bs}{\beta_{\mathrm{s}}}
\def\tran{{\rm T}}

\begin{document}

\title{Scattering from surface fractals in terms of composing mass fractals}

\author{A. Yu. Cherny}
\email[e-mail:~]{cherny@theor.jinr.ru}
\affiliation{Joint Institute for Nuclear Research, Dubna 141980, Russian Federation}
\affiliation{Center for Theoretical Physics of Complex Systems, Institute for Basic Science (IBS), Daejeon 34051, Republic of Korea}

\author{E. M. Anitas}
\affiliation{Joint Institute for Nuclear Research, Dubna 141980, Russian Federation}
\affiliation{Horia Hulubei National Institute of Physics and Nuclear Engineering, RO-077125 Bucharest-Magurele, Romania}

\author{V. A. Osipov}
\affiliation{Joint Institute for Nuclear Research, Dubna 141980, Russian Federation}

\author{A. I. Kuklin}
\affiliation{Joint Institute for Nuclear Research, Dubna 141980, Russian Federation}
\affiliation{Laboratory for Advanced Studies of Membrane Proteins, Moscow Institute of Physics and
Technology, Dolgoprudniy, Russian Federation}

\date{\today}

\begin{abstract}
We argue that a finite iteration of any surface fractal can be composed of mass-fractal iterations of the same fractal dimension. Within this assertion, the scattering amplitude of surface fractal is shown to be a sum of the amplitudes of composing mass fractals. Various approximations for the scattering intensity of surface fractal are considered. It is shown that small-angle scattering (SAS) from a surface fractal can be explained in terms of power-law distribution of sizes of objects composing the fractal (internal polydispersity), provided the distance between objects is much larger than their size for each composing mass fractal. The power-law decay of the scattering intensity $I(q) \propto q^{D_{\mathrm{s}}-6}$, where $2 < D_{\mathrm{s}} < 3$ is the surface fractal dimension of the system, is realized as a non-coherent sum of scattering amplitudes of three-dimensional objects composing the fractal and obeying a power-law distribution $\d N(r) \propto r^{-\tau}\d r$, with $D_{\mathrm{s}}=\tau-1$. The distribution is continuous for random fractals and discrete for deterministic fractals. We suggest a model of surface deterministic fractal, the surface Cantor-like fractal, which is a sum of three-dimensional Cantor dusts at various iterations, and study its scattering properties. The present analysis allows us to extract additional information from SAS data, such us the edges of the fractal region, the fractal iteration number and the scaling factor.
\end{abstract}



\maketitle

\section{\label{sec:Introduction}Introduction}
The small-angle scattering (SAS) of waves (neutrons, X-rays, light) has been proved to be an important non-destructive method of determining the structural properties
at nano and microscales \cite{guinier55:book,svergun87:book,brumberger95:book,lindner02:book}.
These properties are usually obtained from the curve of the elastic cross section per unit volume of the sample (scattering intensity) $I(q) \equiv (1/V^{'}) d \sigma/ d \Omega$ versus the scattering wave vector (momentum) $q=(4\pi /\lambda)\sin \theta$ ($\theta$ is half the scattering angle and $\lambda$ is the wavelength of the incident radiation). The scattering intensity is related to the spatial density-density correlations in the sample by the Fourier transform.

A main indicator of the fractal structure is the power-law dependence of the scattering intensity \cite{bale84,martin87,teixeira88,schmidt91,beaucage96}
\begin{equation}
I(q)\propto q^{-\tau},
\label{eq:sas}
\end{equation}
appearing as a linear dependence on the double logarithm plot within some range in momentum space called the fractal region. This is due to the Hausdorff (fractal) dimension of fractal structures, which is their essential characteristic \cite{mandelbrot83:book, pfeifer83, barnsley88, gouyet96:book}. One can adopt a simple descriptive definition of the Hausdorff dimension $D$ of a set as the exponent in the relation $N\varpropto (L/a)^D$ for $a\to 0$, where $N$ is the minimum number of open sets of diameter $a$ needed to cover the set, and $L$ is the total length of the set. For a `usual' object like ball, the Hausdorff dimensions of volume and surface are equal to 3 and 2, respectively.

Sometimes a succession of simple power-law decays with different exponents can be observed in SAS data, which can be explained by the presence of a few fractal structures  at different scales \cite{beaucage96,chernyJACR14,anitasEPJB14}.

In SAS scattering, one distinguishes between ``mass" and ``surface" fractals \cite{bale84,teixeira88}. The difference can be shown in a simple two-phase geometric configuration, where one phase is a set of dimension $D_\mathrm{m}$ (``mass"), embedded into $d$-dimensional real space, and the other phase is its complement set of dimension $D_\mathrm{p}$ (``pores"). In addition, the boundary between the phases also forms a set of dimension $D_\mathrm{s}$ (``surface"). Then for a
mass fractal, we have $D_\mathrm{s}=D_\mathrm{m}<d$ and $D_\mathrm{p}=d$, while for a surface fractal $D_\mathrm{m}=D_\mathrm{p}=d$ and $d-1<D_\mathrm{s}<d$. Experimentally, the difference between ``mass" and ``surface" fractals \cite{bale84,teixeira88} is revealed through the value of the power-law scattering exponent
\begin{equation}
\tau = \begin{dcases}
   D_{\mathrm{m}}, &\text{for mass fractals},\\
   2d-D_{\mathrm{s}}, &\text{for surface fractals}.
   \end{dcases}
  \label{eq:tau}
\end{equation}
For three-dimensional space ($d=3$), this leads to a simple interpretation of SAS experimental data: if the power-law exponent $\tau <3$, the measured  sample is a mass fractal, while if $3<\tau<4$ then the sample is a surface fractal.

It should be emphasized that the above interpretation of a power-law scattering curve is not rigorous, because the power-law dependence (\ref{eq:tau}) in some region of $q$ can be ``casual". This is a general problem of SAS, since unambiguous interpretation of scattering intensity is hardly possible. Mathematically, in order to restore the spatial dependence of a function, one should know its Fourier transform for \emph{arbitrary} Fourier component $q$. If only a finite range of wave vector is available then this is an ill-posed problem in general.  According to a rule of thumb accepted among experimentalists, if a range, where the power-law dependence is observed, is ``sufficiently large" then the structure is interpreted as a fractal.

For random (statistically self-similar) fractals, one can obtain from SAS data the fractal dimension and, at best, the borders of fractal regions, which give some information about the characteristic lengths of the fractal under investigation (see Sec.~\ref{sec:Theory} below for details). Due to substantial progress in nano-technologies, many deterministic (exactly self-similar) fractal structures were synthesized artificially \cite{mayama06,newkome06, barth07, cerofolini08, polshettiwar09,berenschot13,miloskovskaMACRO14,kajitaPLA14,palmieriAPL14}. As was shown recently \cite{chernyJSI10,chernyJACR10,chernyPRE11}, the scattering intensity of monodisperse deterministic mass fractals shows a generalized power-law decay (maxima and minima superimposed on a simple power-law decay) and contains additional information about the fractals such as the scaling factor, the number of fractal iterations, and the total number of structural units of which the fractal is composed.

Deterministic fractals usually allow analytic solutions for the scattering amplitude and thus give us ``exactly solvable models" for studying the fractal scattering properties. In this paper, we build a model of Cantor-like deterministic surface fractal and investigate its properties. The surface fractal is constructed as a sum of the Cantor dusts with controllable fractal dimension \cite{chernyJSI10,chernyJACR10,chernyPRE11} at various iterations. The construction suggests that in general, \emph{any surface fractal can be represented as a sum of mass fractals}. This is because for mass fractals, the mass and surface dimensions coincide. Therefore, the infinite series of non-overlapping iterations of a mass fractal has the mass dimension $d$, while the surface dimension of the constructed set is equal to the mass fractal dimension. A specific model of such surface fractal is given in Sec.~\ref{subsec:Construction}.

We emphasize a few important issues here. First, a surface fractal can be constructed with subsequent \emph{removal} of mass-fractal iterations from an initial set. For instance, adding mass-fractal iterations to a set is equivalent to subtracting the same iterations from its complement set. However, this does not lead to any problem, because one can always exchange the fractal ``mass" and ``pores" density, thus transforming subtraction into addition. We recall that two complementary sets give the same diffraction pattern (Babinet's principle). Second, the notion of mass fractal should be used here with caution, because the limit of \emph{infinite} iterations might not exist in the rigorous mathematical sense thus giving the empty set in this limit. However, this problem has nothing to do with possible realizations of fractal structures in real materials, because such structures are always \emph{finite}, and, hence, cannot be empty. This means that for a finite iteration of mass fractal, all the scaling fractal properties are confined to a finite range in real space, whether the limit of infinite iterations exists or not. The bigger the iteration number, the longer the fractal range in real space, but the fractal scaling properties within this range would be the same as if the limit of infinite iterations existed.

The construction of a deterministic surface fractal with mass fractals enables us to write down the scattering amplitude of the surface fractal as a sum of the corresponding amplitudes of composing mass fractals. By using this representation, we derive the exponent for the surface fractal intensity [$I(q) \propto q^{D_{\mathrm{s}}-2d}$] from the the scattering intensity for mass fractals [$I(q) \propto V^2 q^{-D_{\mathrm{m}}}$] in various approximations. It is shown that when the distance between objects is much larger than their size for each composing mass fractal, the power-law decay of the scattering intensity of surface fractals is realized as a non-coherent superposition of three-dimensional objects obeying the discrete power-law distribution with the exponent $\tau$, which is shown to be equal $\tau=\ds+1$ with $\ds$ being the surface fractal dimension. The SAS intensity from globular objects obeying the continuous power-law distribution was considered in the paper \cite{schmidt82}. It is shown that the SAS intensity of the discrete distribution has a close analogy to that of the continuous distribution and obeys the generalized power-law decay with the exponent $D_{\mathrm{s}}-2d$.

The paper is organized as follows: in Sec.~\ref{sec:Theory} some important issues concerning SAS are discussed. The section \ref{sec:SAS_MSF} is important for understanding the main ideas of this paper. It shows how the SAS from a surface fractal can be treated in terms of the composing mass fractals within various approximations. The section \ref{sec:dfs} describes the construction of the generalized Cantor surface fractal with controllable dimension, governed by the scaling factor, and the fractal scattering properties are studied.
The internal polydispersity of discrete and continuous types and its role in SAS is considered in Sec.~\ref{sec:rfs}, where we prove that the total surface of objects obeying the power-law distribution with $3<\tau<4$ has the fractal dimension $\ds = \tau-1$. In Conclusion we summarize and discuss the obtained results.

\section{\label{sec:Theory}Theoretical background}

In a very good approximation, the differential cross section of a sample exposed to a beam of neutrons, X-rays or light is given by \cite{guinier55:book,svergun87:book} $\mathrm{d}\sigma/\mathrm{d}\Omega = \left| A(\bm{q}) \right|^{2}$, where $A(\bm{q})\equiv \int_{V'}\rho_{\mathrm{s}}(\bm{r})e^{i\bm{q} \cdot \bm{r}}\mathrm{d}^3 r$ is the total scattering amplitude, $V'$ is the total volume irradiated by the incident beam, and the scattering length density $\rho_{\mathrm{s}}(\bm{r})$ is defined with the help of Dirac's $\delta$ function: $\rho_{\mathrm{s}}(\bm{r})=\sum_{j}{b_{j}\delta(\bm{r}-\bm{r}_{j})}$. Here, $\bm{r}_j$ are the positions of microscopic objects like atoms or nuclei with the scattering lengths $b_{j}$.

Let us consider a sample consisting of rigid \emph{macroscopic} objects of the density $\rho_{\mathrm{m}}$, which are immersed into a solid matrix of density $\rho_{\mathrm{p}}$, and suppose that spatial positions and orientations are uncorrelated (this assumes that the concentration of the objects in the solid matrix is low enough). Then the scattering intensity (differential cross section per unit volume of the sample) can be written as
\begin{equation}
I(q) \equiv \frac{1}{V^{'}}\frac{\mathrm{d}\sigma}{\mathrm{d}\Omega} =  V^{2} \left\langle \left| F(\bm{q}) \right|^{2} \right\rangle,
\label{eq:dcc}
\end{equation}
where $n$ is the concentration of the macroscopic objects in the irradiated volume,  $\Delta \rho = \rho_{\mathrm{m}}-\rho_{\mathrm{p}}$ is the scattering contrast, $V$ is the volume of each object and $F(\bm{q})$ is the normalized scattering amplitude (form factor) of the object
\begin{equation}
F(\bm{q})=\frac{1}{V}\int_{V}e^{-i\bm{q} \cdot \bm{r}}\mathrm{d}\bm{r},
\label{eq:normff}
\end{equation}
obeying the condition $F(0)=1$. Here, the symbol $\left\langle \cdots \right\rangle$ stands for the ensemble averaging over all orientations of the objects. If the probability of any orientation is the same, then it can be calculated by integrating over all directions of the scattering vector $\bm{q}$ \cite{rogachev07}.

It is easy to derive a few useful properties of the form factor (\ref{eq:normff}), which are valid for a particle of arbitrary shape.\\
\noindent \emph{i)} Scaling: if we scale all the lengths of the particle as $l\to \beta l$ then
$F(\bm{q})\to F(\beta\bm{q})$.\\
\noindent \emph{ii)} Translation: if the particle is translated
$\bm{r}\to\bm{r}+\bm{a}$ then $F(\bm{q})\to
F(\bm{q})\exp(-i\bm{q}\cdot\bm{a})$.\\
\noindent \emph{iii)} Rotation: if the particle is rotated with an orthogonal matrix $\bm{r}\to \hat{O}\bm{r}$
then $F(\bm{q})\to F(\hat{O}^\tran \bm{q})$. Recall that the inverse of an orthogonal matrix is equal to the transpose of it $\hat{O}^{-1}=\hat{O}^\tran$, where $(\hat{O}^\tran)_{ij}=\hat{O}_{ji}$.\\
\noindent \emph{iv)} Additivity of the nonnormalized scattering amplitude: if a particle consists of two not overlapping subsets $\mathrm{I}$
and $\mathrm{II}$, then $F(\bm{q}) =\big(V_I F_I(\bm{q}) +V_\mathrm{II}
F_\mathrm{II}(\bm{q})\big)/(V_\mathrm{I}+V_\mathrm{II})$.

\begin{figure}[tb]
\begin{center}
\includegraphics[width=.75\columnwidth]{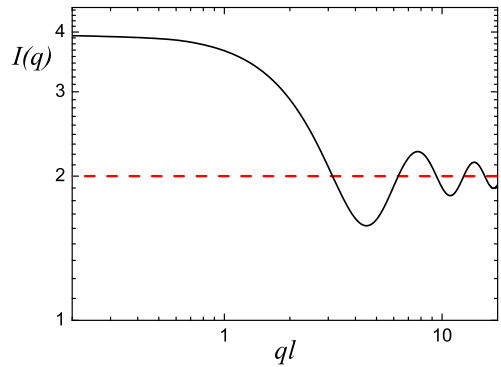}
\caption{\label{fig:2point} (Color online)  The SAS intensity (\ref{2point}) from an ensemble of two point-like objects with unit amplitude, placed rigidly the distance $l$ apart but randomly oriented. One can see the transition from the coherent regime [$I(q)=4$] to the incoherent regime [$I(q)=2$] when $q\gg2\pi/l$: only a very few minima and maxima with decaying amplitudes are quite pronounced. The fast decay of the correlator $\langle e^{i \bm{q}\cdot(\bm{r}_2-\bm{r}_1)}\rangle$ is due to the average over all directions of the scattering vector $\bm{q}$, which is analogous to diffraction with an uncollimated beam in optics \cite{chernyPRE11}: the interference patterns of plane waves, coming from different directions, superimpose upon each other. This results is the same as if the strong spatial incoherence of the incident beam is realized.}
\end{center}
\end{figure}

The average over all directions of the scattering vector $\bm{q}$ in Eq.~(\ref{eq:dcc}) is analogous to diffraction with an uncollimated beam in optics \cite{chernyPRE11}: the interference patterns of plane waves, coming from different directions, superimpose upon each other. This results in strong spatial incoherence: for the subsets
$I$ and $I\!I$, the correlator $\langle F_{I}(\bm{q})F_{I\!I}(\bm{q})\rangle$ decays when $q\gg2\pi/r$, where $r$ is of order of the distance between their centers \cite{chernyPRE11}. This indicates the border between the coherent regime (where the scattering \emph{amplitudes} $V_{I}F_{I}$ and $V_{I\!I}F_{I\!I}$ should be added) and incoherent regime (where the scattering \emph{intensities} $\langle |V_{I}F_{I}|^2\rangle$ and $\langle |V_{I\!I}F_{I\!I}|^2 \rangle$ should be added). This can be illustrated by a simple example of the SAS intensity from two point-like objects, placed rigidly the distance $l$ apart. If each of them has the unit amplitude, the intensity is written as $I(q)=\langle|e^{i \bm{q}\cdot\bm{r}_1}+e^{i \bm{q}\cdot\bm{r}_2}|^2\rangle$, which yields after averaging over the solid angle
\begin{align}\label{2point}
I(q)=2\bigg(1+\frac{\sin ql}{ql}\bigg).
\end{align}
A fast decay of the coherence can be seen from Fig.~\ref{fig:2point} when $ql\gg2\pi$.

For a ``primary" object like a ball or cube of total size $l$, the intensity $\langle |F(\bm{q})|^2 \rangle$ is of order one in the Guinier range $q\lesssim 2\pi/l$ and decays as $1/q^4$ in the Porod range $q\gtrsim 2\pi/l$ \cite{guinier55:book}.

Almost all scattering properties of a complex object can be understood by means of the above simple properties of composing ``primary" objects and transitions from coherent to incoherent scattering regimes. In the next section, we outline and explain some basic properties of mass and surface fractals.

\section{\label{sec:SAS_MSF}General remarks about small-angle scattering from mass and surface fractals}

\subsection{\label{sec:SAS_MF}A mass fractal with a single scale}

The scattering properties of mass fractals with a single scale were studied in detail in the previous publications \cite{chernyJACR10,chernyPRE11}.

\begin{figure}[tb]
\begin{center}
\includegraphics[width=.85\columnwidth]{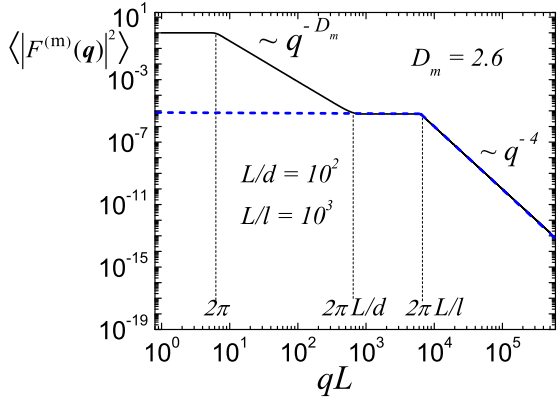}
\caption{\label{fig:MFgen} (Color online) Generic normalized SAS intensity from mass
fractals with a single scale (solid black line). The intensity shows the presence of the four main regimes: Guinier (at small $q$), fractal (at intermediate $q$), plateau (at larger $q$), and Porod (at high $q$). The characteristic lengths $L$, $d$, and $l$ are explained in the text. The blue dashed line shows the approximation of completely uncorrelated primary objects, composing the mass fractal. The scattering intensity of the object (like cube or ball) consists of the Guinier and Porod regions only. Note that a typical experimental SAS tool has the dynamic Q-range $q_{\mathrm max}/q_{\mathrm min}$ about two or three orders, so only a part of the shown curve can be observed in practice.}
\end{center}
\end{figure}

For a mass fractal of the total length $L$, composed of $p$ small ``primary" structural units of size $l$ separated by distances $d$ ($l \lesssim d \ll L$), the normalized form factor can be estimated qualitatively by the formula
\begin{equation}
\big\langle |F^{\mathrm{(m)}}(\bm{q})|^{2} \big\rangle\! \simeq\!
 \begin{dcases}
   1, \!&q \lesssim 2\pi/L,\\
   (qL/2\pi)^{-D_{\mathrm{m}}}, \!&2\pi/L \lesssim q \lesssim 2\pi/d,\\
   (d/L)^{D_{\mathrm{m}}}, \!&2\pi/d \lesssim q \lesssim 2\pi/l,\\
   (d/L)^{D_{\mathrm{m}}}(ql/2\pi)^{-4}, \!&2\pi/l \lesssim q,
 \end{dcases}
\label{eq:ffmassf}
\end{equation}
(see Fig.~\ref{fig:MFgen}). Here $p$ is of the order of $(L/d)^{D_{\mathrm{m}}}$ in accordance with the definition of the fractal dimension.

Such a fractal can be constructed with a simple iteration rule (an example is the Cantor dust considered in Sec.~\ref{subsec:Construction} below): a ``primary" object like a ball or cube or another simple shape generates $k$ objects of the same shape but of the size scaled by the factor $\bs$, which is smaller than one in general. The initial single object (zero iteration) has the size of order $r_0$. Then after $n$ iterations, the total number of the objects is equal to $p=k^n$, and they all are put somehow inside a form of the total size $L$. The distances between the objects and their sizes are of order $d=\bs^n L$ and $l=\bs^n r_0$, respectively. The mass fractal has the Hausdorff dimension $\dm$ obeying the relation \cite{gouyet96:book} $k\bs^{\dm}=1$.

Equation~\eqref{eq:ffmassf} explicitly shows that the SAS intensity of mass fractal is characterized by the four main regions: Guinier at $q \lesssim 2\pi/L$, fractal at $2\pi/L \lesssim q \lesssim 2\pi/d$, a plateau at $2\pi/d \lesssim q \lesssim 2\pi/l$, and Porod regime at $q \gtrsim 2\pi/l$.

We make a few remarks here. First, the intensity in the Guinier range is actually parabolic: $I(q) \simeq I(0)(1-R_{\mathrm{g}}^{2}q^{2}/3)$, where $R_{\mathrm{g}}$ is the radius of gyration. This parabolic behavior of the intensity is ignored in the above estimations for the sake of simplicity. Second, the \emph{mass} fractal region appears due to \emph{spatial correlations} between the composing ``primary" units \cite{chernyJACR10,chernyPRE11}. For this reason, the fractal region of the mass fractal is determined by the maximal and minimal distances between the centers of the structural units. Third, the plateau at $2\pi/d \lesssim q \lesssim 2\pi/l$ in the scattering intensity can be considered as a Guinier region for the primary unit (which is of the same size $l$), because the spatial correlations between different units are not important in this region, and thus the total intensity is equal to $p$ times the intensity of the primary unit (see the discussion in Sec.~\ref{sec:Theory}). For the normalized intensity of primary globular unit of size $l$, one can adopt the Porod-law relation
\begin{equation}
\left\langle \left| F_{0}(\bm{q}) \right|^{2} \right\rangle \simeq
 \begin{dcases}
   1, & q \lesssim 2\pi/l,\\
   (ql/2\pi)^{-4}, &2\pi/l \lesssim q.
 \end{dcases}
\label{eq:IPorod}
\end{equation}
As discussed above, it coincides with the last two rows in Eq.~(\ref{eq:ffmassf}) up to the factor $(d/L)^{D_{\mathrm{m}}}=1/p$, which appears due to the chosen normalization of the total intensity of mass fractal at zero momentum. The latter is equal to $p^2$ times the intensity of the primary unit (the coherent regime). Then neglecting all the spatial correlations \emph{between} the primary objects (units), composing the fractal, yields the scattering intensity shown by the dashed (blue) line in Fig.~\ref{fig:MFgen}. Fourth, the ``pure" power-law functions with different exponents, given by Eq.~(\ref{eq:ffmassf}) and shown in Fig.~\ref{fig:MFgen}, is a simplification of an actual behaviour of the intensity. Actually, there is a complex pattern of maxima and minima superimposed on the power-law decays. However, this pattern is smeared and can disappear completely when the polydispersity is developed~\cite{chernyJACR10,chernyPRE11}.

\subsection{\label{sec:SAS_SF}A surface fractal with a single scale}

In accordance with the statement formulated in the Introduction, \emph{any surface fractal can be constructed as a sum of appropriate mass fractals}. A specific example is given in Sec.~\ref{subsec:Construction} below, see Fig.~\ref{fig:SCF_constr}.

Let us consider the contribution of different mass fractal amplitudes to the total scattering intensity of a surface fractal for a finite iteration $m$. Recall that the non-normalized scattering amplitude is nothing but $V F(\bm{q})$. Because of its additivity [see the property \emph{iv)} in Sec.~\ref{sec:Theory}], one can write the surface fractal amplitude $A_m(\bm{q})$ as a sum of the mass fractal amplitudes~$M_m(\bm{q})$
\begin{equation}\label{SF_ampl_MF}
A_m(\bm{q})=\sum_{n=0}^{m} M_{n}(\bm{q}).
\end{equation}
For simplicity, below in this section we omit the factor $n \left| \Delta \rho \right|^{2}$ in Eq.~(\ref{eq:dcc}) and denote the surface fractal intensity as $I^{\mathrm{(s)}}_m(q)\equiv\langle|A_m(\bm{q})|^2\rangle$. It follows from Eq.~(\ref{SF_ampl_MF}) that the intensity $I^{\mathrm{(s)}}_m(q)$ contains not only the mass fractal intensities $\langle|M_n(\bm{q})|^2\rangle$ but the correlators between the mass-fractal amplitudes
\begin{align}
I^{\mathrm{(s)}}_m&(q)=\sum_{n=0}^{m} \langle|M_{n}(\bm{q})|^2\rangle \nonumber\\
&+\sum_{0\leqslant n<p\leqslant m} \langle M^*_{n}(\bm{q})M_{p}(\bm{q})+M_{n}(\bm{q})M^*_{p}(\bm{q})\rangle.
\label{int_ampl_mass}
\end{align}

\subsubsection{\label{sec:SAS_SF_inMF}The approximation of incoherent mass-fractal amplitudes}

One can neglect completely the non-diagonal (interference) terms in this equation and thus consider the \emph{incoherent sum} of the mass-fractal amplitudes
\begin{align}
I^{\mathrm{(s)}}_m&(q)\simeq\sum_{n=0}^{m} \langle|M_{n}(\bm{q})|^2\rangle.
\label{int_incoh_MA1}
\end{align}
The behaviour of each term in the sum is known from the previous section and shown in Fig.~\ref{fig:MFgen}. Let us show analytically that the surface fractal intensity (\ref{int_incoh_MA1}) obeys approximately the power-law decay with the exponent $6-\ds$, where $\ds=\dm$. For simplicity, we put $d\simeq l\simeq\bs^n L$ in Eq.~(\ref{eq:ffmassf}) thus neglecting the plateau region. We have $\langle|M_{n}(\bm{q})|^2\rangle=V_n^2\langle|F^{\mathrm{(m)}}_{n}(\bm{q})|^2\rangle$ with the volume of the $n$th mass fractal iteration given by $V_n=V_0\bs^{3n}k^n=V_0\bs^{n(3-\dm)}$. Here $V_0$ is the volume of the ``primary" object at zero iteration. If the object is a ball of radius $r_0$ then $V_0=4\pi r_0^3/3$, while for a cube of size $r_0$ it is given by $r_0^3$. With substituting $\langle|M_{n}(\bm{q})|^2\rangle$ into Eq.~(\ref{int_incoh_MA1}), we obtain
\begin{align}
I_m^{\mathrm{(s)}}&(q)=\sum_{n=0}^{m} V^2_0\bs^{2n(3-\dm)}\langle|F^{\mathrm{(m)}}_{n}(\bm{q})|^2\rangle,
\label{int_incoh_MA}
\end{align}
which, in conjunction with Eq.~(\ref{eq:ffmassf}), yields at $q=2\pi/L$
$$
\frac{I_m^{\mathrm{(s)}}(q)}{V_0^2}=\frac{1-\bs^{2(m+1)(3-\dm)}}{1-\bs^{2(3-\dm)}}.
$$
In a similar manner, we obtain at $q=2\pi/(\bs L)$
$$
\frac{I_m^{\mathrm{(s)}}(q)}{V_0^2}=\bs^4+\bs^{6-\dm}\frac{1-\bs^{(2m+1)(3-\dm)}}{1-\bs^{2(3-\dm)}},
$$
and at $q=2\pi/(\bs^2 L)$
\begin{align}
\frac{I_m^{\mathrm{(s)}}(q)}{V_0^2}=\bs^8+\bs^{10-\dm}+\bs^{2(6-\dm)}\frac{1-\bs^{(2m-1)(3-\dm)}}{1-\bs^{2(3-\dm)}}.\nonumber
\end{align}
The above intensities tend to $1/[1-\bs^{2(3-\dm)}]$, $\bs^4+\bs^{6-\dm}/[1-\bs^{2(3-\dm)}]$, and  $\bs^8+\bs^{10-\dm}+\bs^{2(6-\dm)}/[1-\bs^{2(3-\dm)}]$, respectively, for $m\gg 1$. Since $\bs<1$ and $2<\dm=\ds<3$, one can neglect the terms $\bs^4$ and $\bs^8+\bs^{10-\dm}$ in these expressions. This gives us
$$
\frac{I_m^{\mathrm{(s)}}\Big(\frac{2\pi}{L}\Big)}{I_m^{\mathrm{(s)}}\Big(\frac{2\pi}{\bs L}\Big)})\simeq\frac{I_m^{\mathrm{(s)}}\Big(\frac{2\pi}{\bs L}\Big)}{I_m^{\mathrm{(s)}}\Big(\frac{2\pi}{\bs^2 L}\Big)})\simeq\bs^{\ds-6},
$$
that is, the appropriate value of the slope $\ds-6$ on a double logarithmic scale. Similarly, one can consider the intensity at arbitrary wave vectors $q=2\pi/(\bs^n L)$ for $n\leqslant m$.

\subsubsection{\label{sec:SAS_SF_in_units} The approximation of incoherent amplitudes of the primary objects}

One can simplify the above analysis by neglecting the spatial correlations between composing units. We call this approximation \emph{the approximation of incoherent amplitudes of the primary objects} and discuss its applicability below. Then, as discussed in Sec.~\ref{sec:SAS_MF}, one should use the approximation $\langle|F^{\mathrm{(m)}}_{n}(\bm{q})|^2\rangle \simeq
k^{-n}\langle|F_{0}(\bm{q})|^{2}\rangle$ with $l=\bs^n r_0$ in Eq.~(\ref{eq:IPorod}). We denote the intensity of unit at zero iteration as  $I_0(q)\equiv V_0^2\langle|F_{0}(\bm{q})|^{2}\rangle$ with $l=r_0$ and derive from Eq.~(\ref{int_incoh_MA})
\begin{align}
I_m^{\mathrm{(s)}}(q)=\sum_{n=0}^{m} \bs^{n(6-\dm)}I_0(\bs^n q).
\label{eq:intensitysumsquaresv2}
\end{align}

\begin{figure}[tb]
\begin{center}
\includegraphics[width=.82\columnwidth]{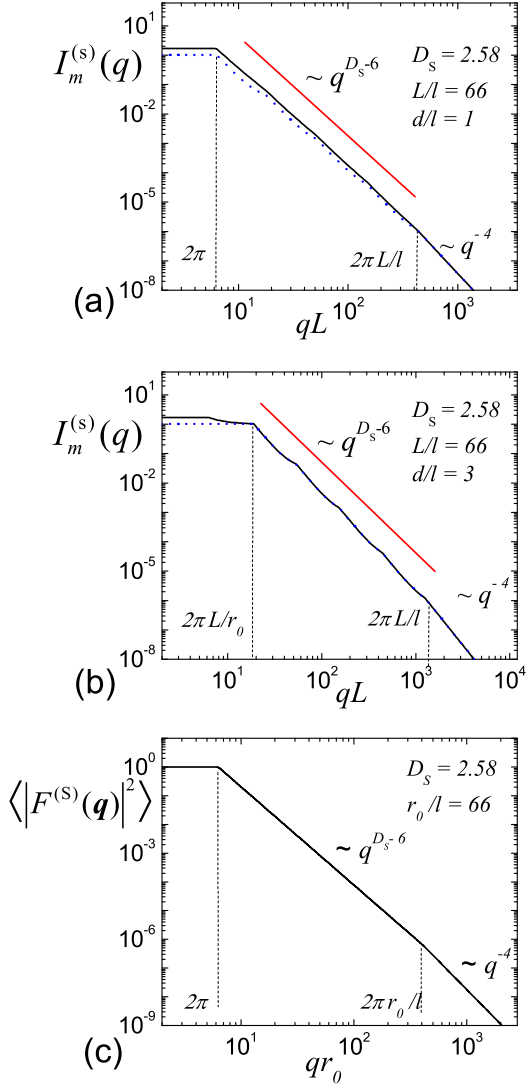}
\caption{\label{fig:SF_sum_MF} (Color online) (a,b) SAS intensity from surface fractal (in units $n |\Delta\rho|^{2} V_{0}^{2}$) versus momentum transfer.
Solid (black) line shows the approximations of incoherent mass-fractal amplitudes (\ref{int_incoh_MA}), and dotted (blue) line shows the approximations of incoherent amplitudes of the primary objects (\ref{eq:intensitysumsquaresv2}) at different values of the control parameters. The intensity represents the three main regimes: Guinier (at small $q$), fractal (at intermediate $q$), and Porod (at high $q$). The bigger the ratio of the distance between primary units $d$ to the their size $l$, the better the approximations of incoherent amplitudes of the primary objects works. (c) Generic normalized SAS intensity from a surface fractal with a single scale. The characteristic lengths $r_0$ and $l$ are of the order of the largest and smallest sizes of the units, respectively. }
\end{center}
\end{figure}

This equation is essential for simple understanding the fractal power-law behaviour of the scattering intensity. The intensity of the unit at zero iteration $I_0(q)$ obeys the Porod law, i.e., $I_0(q)\simeq I_0(0)$ when $q\lesssim 2\pi/r_0$ and starts decreasing as $1/q^4$ when $q\gtrsim 2\pi/r_0$.  Since $\bs^{6-\ds} \ll 1$, the first term in the sum dominates for $q\lesssim 2\pi/r_0$. However, at the point $q\simeq 2\pi/(\bs r_0)$ its contribution becomes about $1/\bs^{4}$ times smaller due to the $1/q^4$ decay, while the second terms is still remains the same. Thus the second term dominates at this point if the surface dimension obeys the inequality $6-\ds<4$. Using the same arguments, we arrive at the conclusion that the $n$th term in Eq.~(\ref{eq:intensitysumsquaresv2}) dominates at the point $q\simeq 2\pi/(\bs^{n-1}r_0)$. Therefore, increasing $q$ by $1/\beta_{\mathrm{s}}$ times leads to decreasing the intensity by $1/\beta_{\mathrm{s}}^{6-D_{\mathrm{s}}}$ times, and the slope of the scattering intensity on a double logarithm scale is $\tau \equiv \log \left( 1/\beta_{\mathrm{s}}^{D_{\mathrm{s}}-6} \right) / \log \left( 1/\beta_{\mathrm{s}} \right) = D_{\mathrm{s}}-6$. We arrive at the power-law behaviour (\ref{eq:sas}), (\ref{eq:tau}) of surface fractal. Note that the inequality $6-\ds<4$ (which follows from $\ds>2$) is crucial in the above consideration. In the case of usual surface dimension $\ds=2$, all the terms in Eq.~(\ref{eq:intensitysumsquaresv2}) decreases as $1/q^4$ and we cannot observe the fractal behaviour of the intensity. The numerical results are shown in Fig.~\ref{fig:SF_sum_MF}a and \ref{fig:SF_sum_MF}b.

The approximation of incoherent amplitudes of the primary objects assumes that the spatial correlations between the primary objects are not important. It happens when $d/l\gg 1$, that is, \emph{the distance between objects is much larger than their size for each mass fractal composing the surface fractal}. The reason is that the correlations between objects' amplitudes decay very fast with growing the distances between their centers (see the discussion in Sec.~\ref{sec:Theory}). Then the surface fractal region lies where the correlations within one mass-fractal iteration have decayed or the contribution of the other mass-fractal intensities are negligibly small.

\begin{figure}[tb]
\begin{center}
\includegraphics[width=.8\columnwidth]{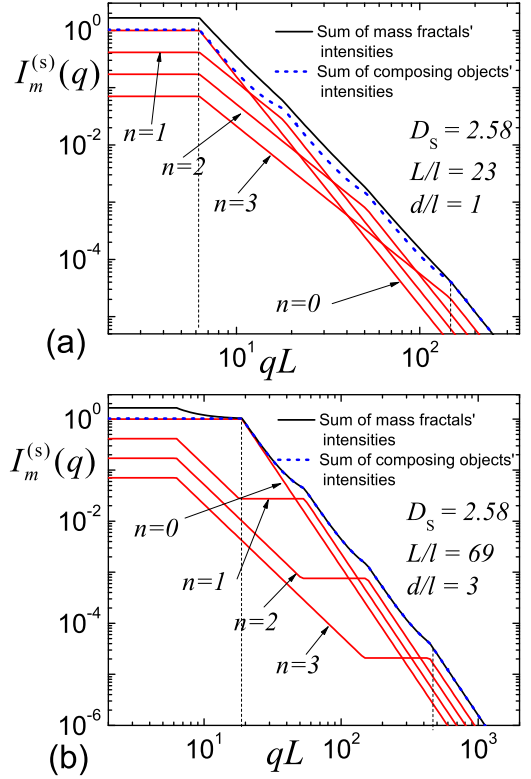}
\caption{\label{fig:SF_sum_MF_det} (Color online) SAS intensity from surface fractal [solid (black) line, in units $n |\Delta\rho|^{2} V_{0}^{2}$) and the SAS intensities of composing mass-fractal iterations [solid (red) lines in the same units] versus momentum transfer.
When the ratio $d/l$ increases, only the Porod regions of the mass fractals contribute to the total intensity of the surface fractal, which means that the approximation of incoherent amplitudes of the primary objects is applicable (see the detailed explanation in the text). The ratio $d/l$ is the same for all mass fractal iterations, and $d/l=L/r_0$.
}
\end{center}
\end{figure}

One can prove this analytically with Eqs.~(\ref{eq:ffmassf}) and (\ref{int_incoh_MA}) in general case (when the plateau presents) by analogy with the derivations in Sec.~\ref{sec:SAS_SF_inMF}. However, one can understand the main features of SAS from the surface fractal directly from Fig.~\ref{fig:SF_sum_MF_det}, which shows contributions of different mass fractals' intensities into the total intensity of the surface fractal.

Indeed, the scattering intensities from mass-fractal iterations [by definition, $\langle|M_{n}(\bm{q})|^2\rangle\equiv V_n^2\langle|F^{\mathrm{(m)}}_{n}(\bm{q})|^2\rangle$] always obey the inequalities $\langle|M_{0}(0)|^2\rangle < \langle|M_{1}(0)|^2\rangle < \ldots < \langle|M_{m}(0)|^2\rangle$. This is because the volume of mass-fractal iterations decreases with its number $n$: $V_n=V_0\bs^{n(3-\dm)}$ (see the discussion in Sec.~\ref{sec:SAS_SF_inMF}), and $\langle|F^{\mathrm{(m)}}_{n}(0)|^2\rangle=1$. The contribution of the zero iteration dominates in its Guinier range $q\lesssim 2\pi/r_0$ because of its largest volume, but for $q\gtrsim 2\pi/r_0$ its intensity decays as $1/q^4$ and can fall off faster than the intensity of the first iteration, which contains the mass fractal range obeying $1/q^{\dm}$ with $\dm<3$, see Fig.~\ref{fig:SF_sum_MF_det}a. Then below the crossover point, the first mass-fractal range contributes substantially to the total surface fractal intensity. In the mass fractal ranges, the correlations between composing units are important, and the approximation of incoherent amplitudes of the primary objects breaks down. On the other hand, if $d/l=r_0/L\gg1$, the plateau is pronounced in each mass fractal region, and we have no intersections between Porod and mass fractal regions of consecutive mass fractal iterations, as one can see from Fig.~\ref{fig:SF_sum_MF_det}b. This means that only the Porod regions contribute to the total intensity of surface fractal, which implies the applicability of the approximation of incoherent amplitudes of the primary objects.


\subsubsection{\label{sec:SAS_SF_in_pairs} The surface fractal intensity in terms of the consecutive mass-fractal iterations}

So far, we consider approximation of incoherent mass-fractal amplitudes (\ref{int_incoh_MA}). However, it might be possible that the spatial distances between  different mass fractal iterations and between composing units within one mass fractal iteration can be of order of their sizes, and we have to take into account the interference  terms in Eq.~(\ref{int_ampl_mass}). This fact does not change the main conclusions of our paper that \emph{the SAS intensity of a surface fractal can always be represented as a sum of intensities of composing mass fractals}. Indeed, considering the correlations between two consecutive mass fractal iterations like $\langle M^*_{0}M_{1}\rangle$, $\langle M^*_{1}M_{2}\rangle$, and so on, and neglecting the other correlations, we obtain from Eq.~(\ref{int_ampl_mass})
\begin{align}
I^{\mathrm{(s)}}_{m+1}(q)\simeq\sum_{n=0}^{m} \langle|M_{n}(\bm{q})+M_{n+1}(\bm{q})|^2\rangle -\sum_{n=1}^{m} \langle|M_{n}(\bm{q})|^2\rangle.
\label{int_double_mass}
\end{align}
The first sum in the approximation (\ref{int_double_mass}) is \emph{incoherent sum} of intensities of \emph{pairs} of consecutive amplitudes. Formally, the sum of two consecutive mass-fractal iterations is nothing else but a mass fractal with the same single scale. It can be considered as a mass fractal with complex composing units. Then, in accordance with the above discussions, its SAS intensity behaves like a mass fractal with the power-law decay $I^{\mathrm{(m)}}_n(q)\sim q^{-\dm}$. Applying the same arguments as in Sec.~\ref{sec:SAS_SF_inMF} yields the power-law decay of the intensity (\ref{int_double_mass}): $I^{\mathrm{(s)}}_m(q)\sim q^{D_\mathrm{s}-6}$ at $\dm=\ds$. In the same manner as in Sec.~\ref{sec:SAS_SF_in_units}, we obtain that Eq.~(\ref{int_double_mass}) leads to the approximation of incoherent amplitudes of the primary objects when $d\gg l$.

By analogy with the pair consecutive amplitudes, one can further improve the approximation (\ref{int_double_mass}) for the SAS intensity by including the triple consecutive amplitudes $\langle|M_{n}+M_{n+1}+M_{n+2}|^2\rangle$.

The approximations for the surface fractal amplitude are considered in Sec.~\ref{analysis} below.

\subsubsection{\label{sec:SAS_SF_generic} The generic scattering intensity from a surface fractal with a single scale}

For a surface fractal composed of ``primary" units,
the qualitative formula for the normalized SAS intensity takes the form
\begin{equation}
\big\langle | F^{\mathrm{(s)}}(\bm{q}) |^{2} \big\rangle \!\simeq\!
 \begin{dcases}
   1, \!\!&q \lesssim 2\pi/r_0,\\
   (qr_0/2\pi)^{D_{\mathrm{s}}-6}, \!\!&2\pi/r_0 \lesssim q \lesssim 2\pi/l,\\
   (r_0/l)^{D_{\mathrm{s}}-6}(ql/2\pi)^{-4}, \!\!&q \gtrsim 2\pi/l,
 \end{dcases}
\label{eq:ffsurfacef}
\end{equation}
(see Fig.~\ref{fig:SF_sum_MF}c), and in this case $r_0$ and $l$ are of the order of the largest and smallest sizes of the units, respectively. This approximation always reproduce correctly the \emph{borders} of the fractal region for a surface fractal and the rough structure of the scattering intensity.

\section{\label{sec:dfs}Deterministic surface fractals}

\subsection{\label{subsec:Construction}Construction and properties}

\begin{figure}[tb]
\begin{center}
\includegraphics[width=\columnwidth]{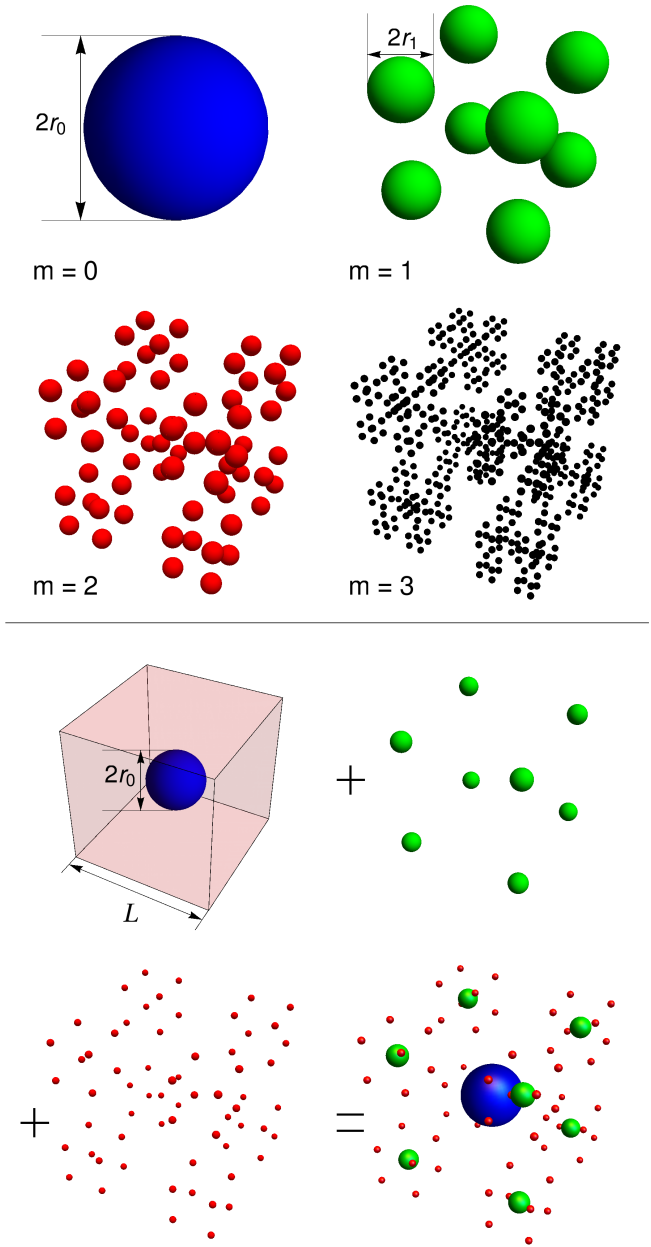}
\end{center}\caption{
(Color online) Upper panel: The initiator (m=0) and first three iterations of the \emph{mass} generalized Cantor fractal (Cantor dust). Each ball of radius $r_m$ generates $k=8$ balls of radius $r_{m+1}=\beta_{\mathrm{s}}r_m$ at each subsequent iteration; Lower panel: the second iteration of the Cantor-like \emph{surface} fractal that is a sum of the mass fractals of zeroth, first, and second iterations.}
\label{fig:SCF_constr}
\end{figure}

The Cantor-like surface fractal is constructed as a sum of mass generalized Cantor fractals (GCF), which are suggested and discussed in detail in Refs.~\cite{chernyJSI10,chernyJACR10,chernyPRE11}. The GCF is also called Cantor dust. Let us recall the construction algorithm for the GCF. We start with a cube of edge ${L}$ and choose a Cartesian system of coordinates with the origin in the cube center, and the axes parallel to the cube edges. The zeroth iteration (called initiator) is a ball of radius $r_{0}$ in the origin.
The iteration rule (generator) is to replace the ball with $k$ smaller balls ($k=8$) of radius $r_{1}=\beta_{\mathrm{s}}r_{0}$, where the parameter $\beta_{\mathrm{s}}$, called scaling factor, obeys the condition $0 <\beta_{\mathrm{s}} <1/2$. The centers of the eight balls of radius $r_{1}$ are shifted from the origin by the eight vectors
\begin{equation}
\bm{a}_{j}=\frac{1-\beta_\mathrm{s}}{2}{L}\left\{ \pm 1, \pm 1, \pm 1\right\}
\label{eq:shifts}
\end{equation}
with all the combinations of the signs. The next iterations are obtained by performing an analogous operation to each of $k$ balls of radius $r_{1}$, and so on
(see Fig.~\ref{fig:SCF_constr}). The fractal dimension of the Cantor dust (mass Cantor fractal) is given by \cite{chernyJACR10}
\begin{equation}\label{dimmass}
D_{\mathrm{m}} =-\ln k/\ln\beta_{\mathrm{s}}
\end{equation}
with $k=8$ for the Cantor dust in three dimensions. It lies within $0<D_{\mathrm{m}}<3$. We emphasis that the  \emph{Hausdorff (fractal)} dimension of the total \emph{volume} of the balls coincides with that of the total \emph{surface} of the spheres in the limit $m\to\infty$. This is a seemingly paradoxical conclusion resulted from the infinite mathematical procedure $m\to\infty$. The coincidence of the volume dimension and surface dimensions in the GCF is a generic characteristic of mass fractal (see Introduction).

The $m$-th iteration of the three-dimensional Cantor-like surface fractal is built as a \emph{sum} of the Cantor dusts of iterations from zero to $m$, see Fig.~\ref{fig:SCF_constr}. In order to avoid the overlapping between the different iterations of the Cantor dust, the initial radius should be restricted: $r_0\leqslant {L}(1-2\beta_{\mathrm{s}})/2$.
By the construction, the initial length ${L}$ is nothing else but the size of the surface fractal if $m$ is big enough. The essential difference between the Cantor mass and surface fractals is that, at a given iteration, the mass fractal consists of subunits with the same size, while the surface fractal consists of subunits with different sizes, obeying the discrete power-law distribution.
The difference is apparent from Fig.~\ref{fig:SCF_constr}.

At the $m$-th iteration, the three-dimensional Cantor-like surface fractal is composed of $N_m=1+k+k^2+\cdots+k^m$ balls
\begin{equation}\label{Nm}
N_m=(k^{m+1}-1)/(k-1)
\end{equation}
(with $k=8$), whose radii and volumes are distributed in the following way. One ball of radius $r_0$ has volume $4\pi r_{0}^{3}/3$, $k$ balls of radius $r_{1}=\beta_{\mathrm{s}}r_{0}$ have the volume $k 4\pi r_{1}^{3}/3$, $k^{2}$ balls of radius $r_{2}=\beta_{\mathrm{s}}^{2}r_{0}$ have the volume $k^{2}4\pi r_{2}^{3}/3$), and so on. Then, the total volume of surface fractal at $m$-th iteration is given by
\begin{equation}\label{totVSF}
V_{m}=V_0\frac{1-(k\beta_{\mathrm{s}}^3)^{m+1}}{1-k\beta_{\mathrm{s}}^3}
\end{equation}
with the volume of zero iteration $V_0=4\pi r_{0}^{3}/3$. Because of the inequality $k\beta_{\mathrm{s}}^3<1$, the total volume (\ref{totVSF}) is finite in the limit $m\to\infty$, and then the Hausdorff dimension of the fractal \emph{volume} is equal to 3.

The contribution of the initiator ($m=0$) to the Hausdorff dimension of the total \emph{surface} of the Cantor-like fractal is obviously equal to 2, which yields the lower limit for the surface dimension, while the contribution of the $m$-th mass iteration for $m\to \infty$ is given by the fractal dimension (\ref{dimmass}). Then we arrive at the the Hausdorff (fractal) dimension of the total \emph{surface} of the Cantor-like fractal
\begin{equation}
D_{\mathrm{s}} =\begin{dcases}
   2, &\text{for}\ 0<\beta_{\mathrm{s}}\leqslant 1/\sqrt{k},\\
   -\ln k/\ln\beta_{\mathrm{s}}, &\text{for}\ 1/\sqrt{k}\leqslant\beta_{\mathrm{s}}<1/2.
   \end{dcases}
\label{eq:sfddeter}
\end{equation}
The threshold  value $\beta_{\mathrm{s}}=1/\sqrt{k}$ corresponds to $D_{\mathrm{m}}=2$ in Eq.~(\ref{dimmass}), which yields $\beta_{\mathrm{s}}=1/(2\sqrt{2})=0.353\ldots$ for $k=8$. When the scaling factor $\beta_{\mathrm{s}}$ is smaller than this value, the total surface of the fractal is finite even in the limit $m\to\infty$. As expected \cite{bale84,martin87,schmidt91}, the surface Hausdorff dimension satisfies the condition $2 \leqslant  D_{\mathrm{s}} <3$.

\subsection{\label{subsec:monoformfactor}Monodisperse fractal form factor}

At $n$-th iteration the mass GCF is composed of balls of the same size $\beta_{\mathrm{s}}^{n}r_0$. The normalized scattering amplitude is known analytically~\cite{chernyJSI10,chernyJACR10}
\begin{equation}
F_{n}^{\mathrm{(m)}}(\bm{q})=F_{0}(\beta_{\mathrm{s}}^{n} q r_{0})G_{1}(\bm{q})G_{1}(\beta_{\mathrm{s}}\bm{q}) \cdots G_{1}(\beta_{\mathrm{s}}^{n-1}\!\bm{q}),
\label{eq:massformfactor}
\end{equation}
where $G_{1}(\bm{q}) \equiv \cos(uq_{x})\cos(uq_{y})\cos(uq_{z})$ is the generative function depending on the relative positions of the balls inside the first iteration of the fractal. Here
\begin{equation}
F_{0}(z) = 3(\sin z - z \cos z)/z^{3},
\label{eq:ballformfactor}
\end{equation}
is the form factor of ball of unit radius, and $u \equiv {L}(1-\beta_{\mathrm{s}})/2$. One can put by definition for the zeroth iteration $F_{0}^{\mathrm{(m)}}(\bm{q})=F_{0}(q r_{0})$.

The surface fractal, by its intrinsic construction (see the previous section), is the sum of mass GCF at various iterations, and, hence, we should add the amplitudes of the mass fractal iterations $V_0(k\beta_{\mathrm{s}}^3)^{n}F_{n}^{\mathrm{(m)}}(\bm{q})$ and normalize the result to one at $q=0$
\begin{equation}
F_{m}^{\mathrm{(s)}}(\bm{q})= \frac{1-k\beta_{\mathrm{s}}^3}{1-(k\beta_{\mathrm{s}}^3)^{m+1}}\sum_{n=0}^{m}(k\beta_{\mathrm{s}}^3)^{n}F_{n}^{\mathrm{(m)}}(\bm{q}),
\label{eq:surfaceformfactor}
\end{equation}
where the normalization condition $F_{m}^{\mathrm{(s)}}(0)=1$ is satisfied. Then the scattering intensity is calculated with Eq.~(\ref{eq:dcc})
\begin{equation}
I_{m}^{\mathrm{(s)}}(q)= I_{m}^{\mathrm{(s)}}(0)\Big\langle \big|F_{m}^{\mathrm{(s)}}(\bm{q})\big|^{2}\Big\rangle
\label{eq:finalintensity}
\end{equation}
with $I_{m}^{\mathrm{(s)}}(0)=n \left| \Delta \rho \right|^{2} V_m^{2}$, where $V_m$ is given by Eq.~(\ref{totVSF}).

The radius of gyration $R_{\mathrm{g}}$ is related to the expansion of the scattering intensity for $q\to 0$~\cite{svergun87:book}
\begin{equation}
I(q) = I(0)(1-q^{2}R_\mathrm{g}^{2}/d+\cdots)
\label{eq:guinierregion}
\end{equation}
with $d=3$ for three-dimensional space. The calculations of the fractal radius of gyration can be simplified, since the expansion of form factor (\ref{eq:massformfactor}) is radially symmetric up to quadratic terms in $\bm{q}$ due to the cube rotational symmetry. This implies that the total form factor of the  surface fractal (\ref{eq:surfaceformfactor}) has the same symmetry as well, which leads to $F_{m}^{\mathrm{(s)}}(\bm{q}) = 1-q^{2}R_\mathrm{g}^{2}/6+\cdots$. Expanding Eq.~(\ref{eq:massformfactor}), substituting the result into Eq.~(\ref{eq:surfaceformfactor}), and combining the terms proportional to $q^2$ yield
\begin{equation}
R_{\mathrm{g}}=\left(\frac{3}{5}r_{0}^{2}\mu +3\frac{1-\beta_{\mathrm{s}}}{1+\beta_{\mathrm{s}}}{L}^2\nu\right)^{1/2},
\label{eq:sfradiusofgyration}
\end{equation}
where the dimensionless parameters $\mu$ and $\nu$ are given by
\[\mu\equiv\frac{1-x}{1-y}\frac{1-y^{m+1}}{1-x^{m+1}},\ \nu\equiv x\frac{1-x^m}{1-x^{m+1}}-y\frac{1-x}{1-y}\frac{1-y^{m}}{1-x^{m+1}}
\]
with $x\equiv k\beta_{\mathrm{s}}^3$ and $y\equiv k\beta_{\mathrm{s}}^5$. The radius of gyration of the Cantor surface fractal takes a simple form when  $m\to\infty$
\begin{equation}
R_{\mathrm{g}}=\left(\frac{3}{5}\frac{1-k\beta_{\mathrm{s}}^3}{1-k\beta_{\mathrm{s}}^5}r_{0}^{2} +
3\frac{(1-\beta_{\mathrm{s}})^2}{1-k\beta_{\mathrm{s}}^5}k\beta_{\mathrm{s}}^3{L}^2\right)^{1/2}.
\label{eq:sfradiusofgyrationinf}
\end{equation}

\subsection{\label{subsec:polyformfactor}Polydisperse fractal form factor}
In most cases, a real system consists of fractals of various sizes and forms (polydispersity). We can model polydispersity by considering an ensemble of GCF with different lengths $l$ of the initial cube taken at random (that is, $l$ is here the length of the initial cube and the ratio $l/r_0$ is held constant over the ensemble, see Sec.~\ref{subsec:Construction}). Note that in the previous sections, we denote the length of the initial cube ${L}$, while in the presence of polydispersity, ${L}$ is the mean value of the cube length over the ensemble.

The distribution function $D_{N}(l)$ of the fractal sizes is defined in such a way that $D_{N}(l)\d l$ gives the probability of finding a fractal whose size falls within the range ($l, l+\d l$). We consider here quite common log-normal distribution
\begin{equation}
D_{N}(l)=\frac{1}{\sigma l(2\pi)^{1/2}}\exp\left(-\frac{[\log(l/{L})+\sigma^{2}/2]^{2}}{2\sigma^{2}}\right),
\label{eq:log-normal}
\end{equation}
where $\sigma=[\log(1+\sigma_{\mathrm{r}}^{2})]^{1/2}$. The quantities ${L}$ and $\sigma_{\mathrm{r}}$ are the mean length and its coefficient of variation (that is, the ratio of the standard deviation of the length to the mean length),  called also relative variance
\begin{equation}
{L}\equiv \left\langle l \right\rangle_{D},~~~\sigma_{\mathrm{r}} \equiv (\left\langle l^{2} \right\rangle_{D}-{L}^{2})^{1/2}/{L},
\label{eq:meanandvar}
\end{equation}
where $\left\langle \cdots \right\rangle_{D} \equiv \int_{0}^{\infty} \cdots D_{N}(l)\mathrm{d}l$. Therefore, by using Eqs.~\eqref{eq:dcc} and~\eqref{eq:log-normal} the polydisperse intensity becomes
\begin{equation}
I^{\mathrm{(s)}}_{m}(q)=n\left| \Delta \rho \right|^{2}\int_{0}^{\infty}{ \left\langle \left| F_{m}^{\mathrm{(s)}}(\bm{q}) \right|^{2} \right\rangle V_{m}^{2}(l)D_{N}(l)\mathrm{d}l},
\label{eq:polyint}
\end{equation}
where the amplitude is given by Eq.~\eqref{eq:surfaceformfactor}.

\subsection{Analysis of the main regions in the scattering intensity}
\label{analysis}

The numerical results for the SAS intensities of the first three iterations of the surface Cantor fractal are shown in Fig.~\ref{fig:SCF_int}. One can clearly distinguish four main subsequent regions: the Guinier, intermediate, surface fractal, and Porod regions.

\subsubsection{The Guinier and intermediate regions}
\label{GIregion}

In the \emph{Guinier region} $q\lesssim 2\pi/{L}$, we deal with completely coherent scattering of all structural units with zero phase difference. Thus, the spatial correlations at the distance of order of the overall fractal size ${L}$ are important.

In the \emph{intermediate region}, we observe a quite complicated interference (\ref{eq:finalintensity}) of the scattering amplitudes of mass Cantor fractals  (\ref{eq:massformfactor}) composing the surface Cantor fractal. The scattering from Cantor-like mass fractals was studied in detail in Refs.~\cite{chernyJSI10,chernyJACR10,chernyPRE11}.

The correlations of amplitudes of structural fractal units decay subsequently with increasing $q$. Thus, the correlations \emph{between the amplitudes of different mass fractal iterations} decay (that is  $\langle F_{n}^{\mathrm{(m)}}(\bm{q})F_{j}^{\mathrm{(m)}}(\bm{q})\rangle \simeq 0$ for $n\neq j$) when $q\gtrsim 2\pi/r_{nj}$ with $r_{nj}$ being a typical distance between balls in the $n$th and $j$th mass fractal iterations. Then we derive from Eqs.~(\ref{eq:surfaceformfactor}) and (\ref{eq:finalintensity})
\begin{align}
I_{m}^{\mathrm{(s)}}&(q)/I_{m}^{\mathrm{(s)}}(0)=\langle |F_{m}^{\mathrm{(s)}}(\bm{q})|^2\rangle \nonumber\\
&\simeq \frac{(1-k\beta_{\mathrm{s}}^3)^2}{\big(1-(k\beta_{\mathrm{s}}^3)^{m+1}\big)^2}\sum_{n=0}^{m}(k\beta_{\mathrm{s}}^3)^{2n}\langle |F_{n}^{\mathrm{(m)}}(\bm{q})|^2\rangle,
\label{eq:SF_incoh_mass}
\end{align}
where $k=8$.

Further, for $n$th mass fractal iteration, the spatial correlations between the ball positions become immaterial at the upper border of \emph{mass} fractal range $q  \simeq 4\pi/ [(1-\beta_\mathrm{s})\beta_\mathrm{s}^{n-1}{L}]$ and higher due to transition from to the incoherent scattering regime, where we have $\langle |F_{n}^{\mathrm{(m)}}(\bm{q})|^2\rangle\simeq F_0^2(\bs^n q)/k^n$, see Refs.~\cite{chernyJACR10,chernyPRE11}. Therefore, when the correlations \emph{between the amplitudes of all balls} composing the surface fractal are negligible, we obtain from Eq.~(\ref{eq:SF_incoh_mass})
\begin{align}
I_{m}^{\mathrm{(s)}}&(q)/I_{m}^{\mathrm{(s)}}(0)=\langle |F_{m}^{\mathrm{(s)}}(\bm{q})|^2\rangle \nonumber\\
&\simeq \frac{(1-k\beta_{\mathrm{s}}^3)^2}{\big(1-(k\beta_{\mathrm{s}}^3)^{m+1}\big)^2}\sum_{n=0}^{m}k^{n}\beta_{\mathrm{s}}^{6n}F_{0}^2(\beta_{\mathrm{s}}^{n} q r_{0}).
\label{eq:SF_incoh_sph}
\end{align}

\begin{figure}[tb]
\begin{center}
\includegraphics[width=\columnwidth]{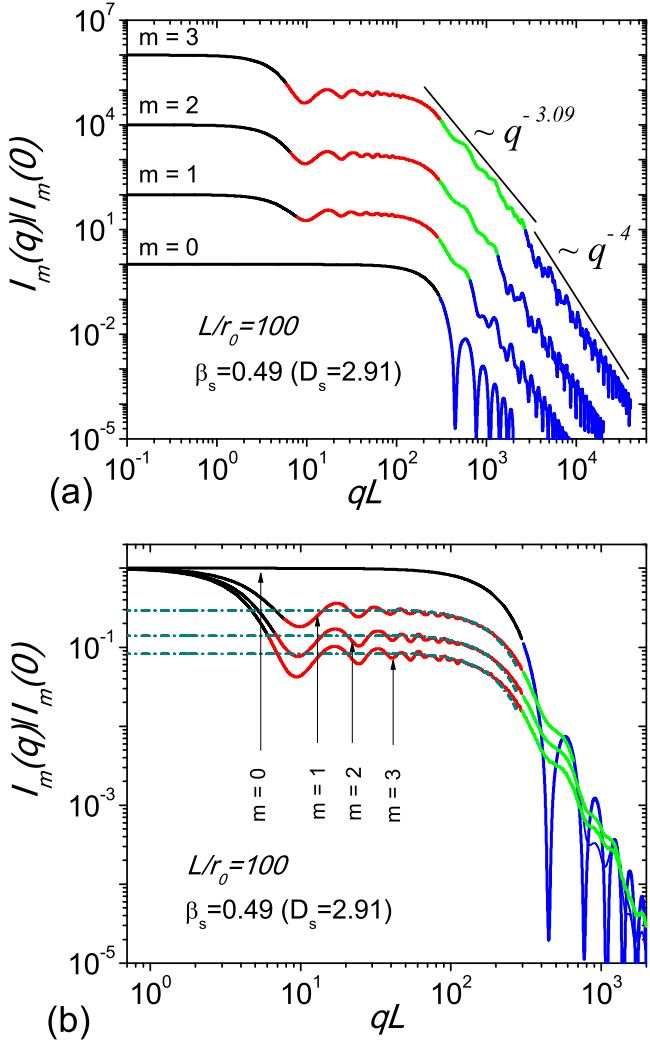}
\end{center}\caption{(Color online) Scattering intensity (\ref{eq:finalintensity}), normalized to one at $q=0$, for the first three iterations of the monodisperse surface fractal versus the wave vector (in units of the inverse total fractal size). (a) Scattering curve for the $m$th iteration is scaled up for clarity by the factor $10^{2m}$. The Guinier, intermediate, fractal, and Porod regions are shown in black, red, green, and blue, respectively. (b) Asymptotes of the plateau (\ref{eq:intensityatzero}) are indicated in ash-dot cyan.}
\label{fig:SCF_int}
\end{figure}

Besides, each ball of radius $r_0\bs^{n}$ behaves as a point-like object with $F(q) \simeq 1$ unless the wave vector gets larger than about $\pi/(r_0\bs^{n})$, see the discussion in Sec.~\ref{sec:Theory}. This means that we observe an interference pattern of the point-like objects with the amplitudes proportional to their volumes $V_0\bs^{3n}$ (here $V_0 = 4\pi r_{0}^{3}/3$)  up to $q {L}\lesssim 2\pi {L}/r_0=100\pi$ at the chosen values of control parameters in Fig.~\ref{fig:SCF_int}.

We clearly see the second plateau where all the correlations between the ball amplitudes have decayed but the balls still scatter as point-like objects. Replacing $F_0$ by one and summing the remaining terms in Eq.~(\ref{eq:SF_incoh_sph}) yield the asymptotic value $I^{\mathrm{as}}_{m}$ of the second plateau
\begin{equation}
I^{\mathrm{as}}_{m}/I_{m}^{\mathrm{(s)}}(0)\simeq \frac{(1-k\beta_{\mathrm{s}}^3)^2}{\big(1-(k\beta_{\mathrm{s}}^3)^{m+1}\big)^2}
\frac{1-(k\beta_{\mathrm{s}}^6)^{m+1}}{1-k\beta_{\mathrm{s}}^6}.
\label{eq:intensityatzero}
\end{equation}
Note that the second plateau can be considered as the Guinier region for a surface fractal composed of spatially uncorrelated objects, see Eq.~(\ref{eq:ffsurfacef}).

We emphasize the following point. The surface fractal is composed of the mass fractals. In spite of this fact, only the scattering pattern from the first mass-fractal iteration manifests itself in the intermediate region shown in Fig~\ref{fig:SCF_int} at the chosen values of the control parameters. If, however, the Cantor surface fractal construction starts from the $n$th Cantor mass fractal with $n\gg1$, one can observe a clearly pronounced \emph{mass fractal regime}. This is a specific feature of the surface fractal construction, which is not related to the surface fractal region, and we will discuss this property elsewhere \cite{tbp}. Instead, in this paper we focus on the next \emph{surface fractal region} with a complex pattern of maxima and minima superimposed on a power-law decay $I(q)\sim 1/q^{6-D_{\textrm{s}}}$ (generalized power-law decay).

\begin{figure}[tb]
\begin{center}
\includegraphics[width=\columnwidth]{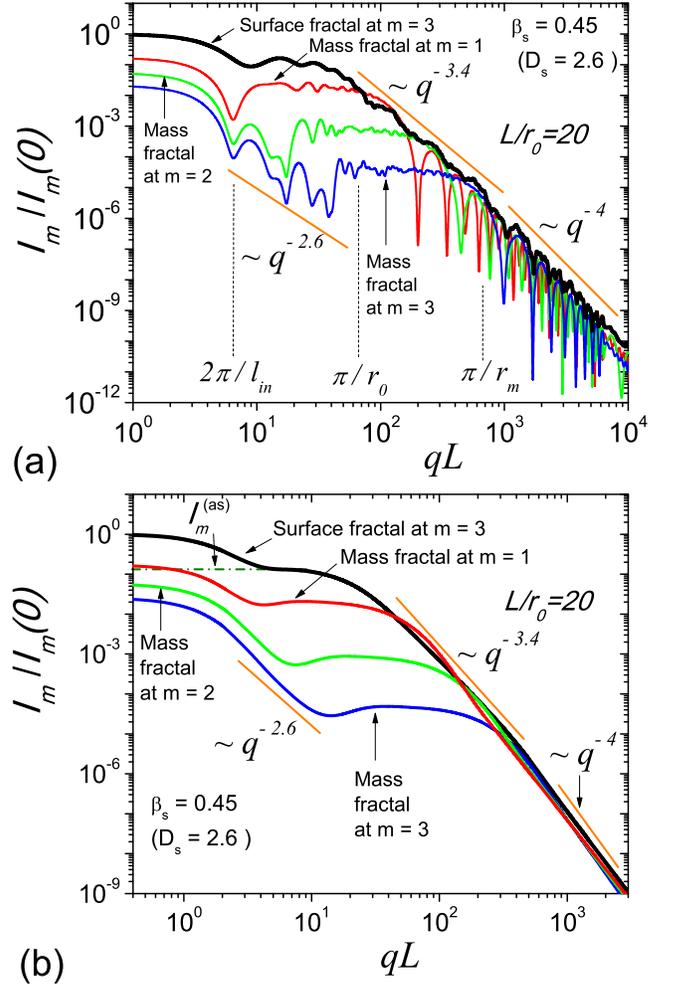}
\end{center}\caption{(Color online) Scattering from entire surface fractal [Eq.~\eqref{eq:finalintensity}] and the separate contributions of the mass fractal iterations composing the surface fractal. Here $r_m=\bs^m r_0$ is radius of the balls for the $m$th iteration.
(a) Monodisperse scattering. (b) Polydisperse scattering with relative variance $\sigma_{\mathrm{r}}=0.4$ (compare with the model curves shown in Fig.~\ref{fig:SF_sum_MF_det}b) }
\label{fig:SCF_int_sum}
\end{figure}

\subsubsection{The surface fractal and Porod regions}
\label{SFPregion}

If the ratio $d/l=L/r_0$ is chosen to be large enough, the fractal region of a surface fractal arises as a result of \emph{incoherent} diffraction of all units composing the fractal (see the discussion in Sec.~\ref{sec:SAS_SF_in_units}). This means that we should \emph{add up intensities} of the fractal units together but not their amplitudes. Then the scattering intensity can be easily calculated in the fractal region, once the fractal structure is known. For the Cantor surface fractal (see Sec.~\ref{subsec:Construction}), we have, first, the contribution of the central ball (the first mass fractal iteration), see Sec.~\ref{sec:Theory},  $I_0(q)\equiv n |\Delta\rho|^{2} V_{0}^{2}F_0^2(q r_0)$ with $V_0=4\pi r_0^3/3$ and $F_0$ being the ball volume and its form factor (\ref{eq:ballformfactor}), respectively. Second, the contribution of the first mass fractal iteration is $k\bs^6 I_0(\bs q)$ (because it consists of $k=8$ balls with radii $\bs r_0$), and so on. Repeating all the arguments of Sec.~\ref{sec:SAS_SF_in_units}, we explain the exponent $\ds-6$ in the fractal region of the surface fractal.

For high wave vectors $q\gtrsim \pi/r_m$, we have \emph{the Porod region}, which is determined by the size of the smallest fractal subunits, balls of radius $r_m=\bs^m r_0$. In the Porod region, the scattering intensity resembles the intensity of the initiator (a ball in our case), obeying the Porod law $1/q^4$.

\begin{figure}[tb]
\centerline{\includegraphics[width=\columnwidth]{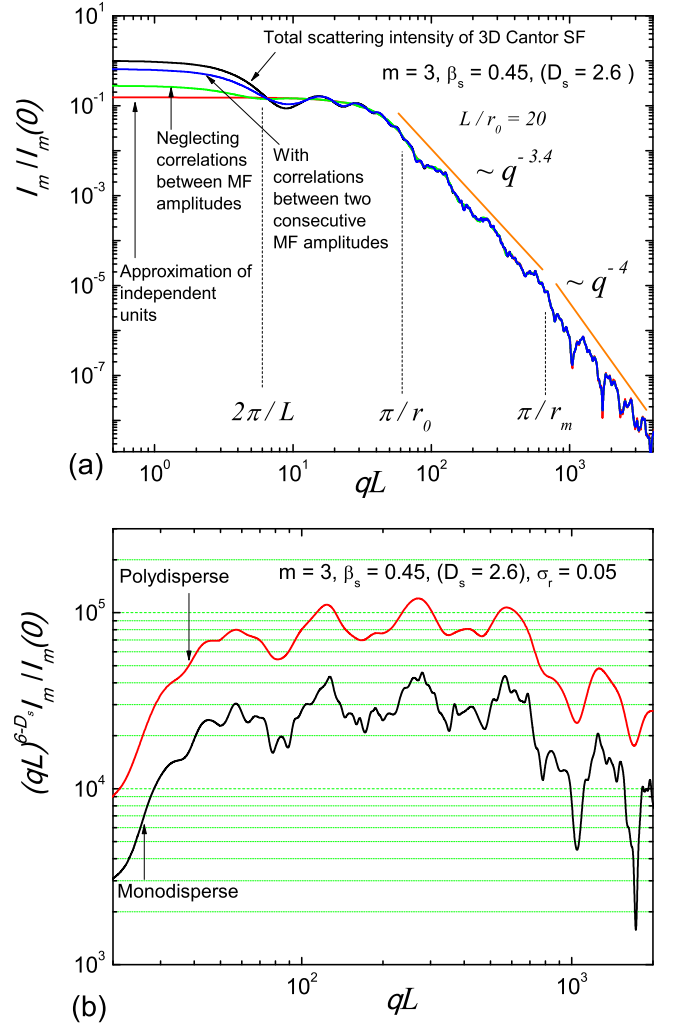}}
\caption{(Color online) Scattering intensity from surface fractals. (a) Exact total scattering intensity (\ref{eq:finalintensity}), the approximation  (\ref{eq:SF_incoh_mass}) neglecting the correlations between mass fractal amplitudes, and completely incoherent sum of intensities of the balls (\ref{eq:SF_incoh_sph}) are shown in black, red, and green, respectively. The fine structure of the intensity is approximated fairly well by the incoherent sum of intensities of the balls in the fractal region, since the ratio $L/r_0=d/l$ is large. (b) The scaled scattering intensity $(qL)^{6-D_{\mathrm{s}}}I(q)$, shown in black, is a log-periodic function with the factor $1/\bs$. Polydispersity (red curve, scaled up by the factor $3$ to facilitate visualization) smoothes the minima and maxima spreading. The relative variance of polydispersity $\sigma_{\mathrm{r}}$ is equal to $0.05$.}
\label{fig:CSF_incoher_sum}
\end{figure}

Figures \ref{fig:SCF_int_sum} and \ref{fig:CSF_incoher_sum}a illustrates that the scattering intensity of a surface fractal in the fractal range is actually realized as a \textit{non-coherent sum} of intensities of a system of balls. One can see from Fig.~\ref{fig:CSF_incoher_sum}a that in the fractal region $\pi/r_0 \lesssim q \lesssim \pi/r_{m}$, we have a very good coincidence between exact formula (\ref{eq:finalintensity}), the approximation  (\ref{eq:SF_incoh_mass}) neglecting the correlations between mass fractal amplitudes, and completely incoherent sum of intensities of the balls (\ref{eq:SF_incoh_sph}), which are discussed in detail in Sec.~\ref{GIregion}.

In order to observe deviations form the surface fractal power-law $1/q^{6-D_{\mathrm{s}}}$, one can scale out it and thus depict $q^{6-D_{\mathrm{s}}}I(q)$ as a function of $q$ in a log-scale, see Fig.~\ref{fig:CSF_incoher_sum}b. The minima and maxima exhibit an approximate log-periodicity  with the scale factor ${1/\beta_{\mathrm{s}}}$. This result has analogy with deterministic mass fractals \cite{chernyPRE11}, but its nature is different. Indeed, the log-periodicity in mass fractals arises from the self-similarity of \emph{distances} between the structural units, while the log-periodicity in surface fractals arises from the self-similarity of \emph{sizes} of the structural units. As usual, polydispersity smoothes the minima and maxima spreading, which can can have a dramatic effect on possible experimental observations. Nevertheless, the effect still appears when polydispersity is not high, and the log-periodicity allows us to extract information about the scale factor $1/\bs$ of deterministic surface fractals from SAS intensity obtained experimentally.

\begin{figure}[tb]
\centerline{\includegraphics[width=\columnwidth]{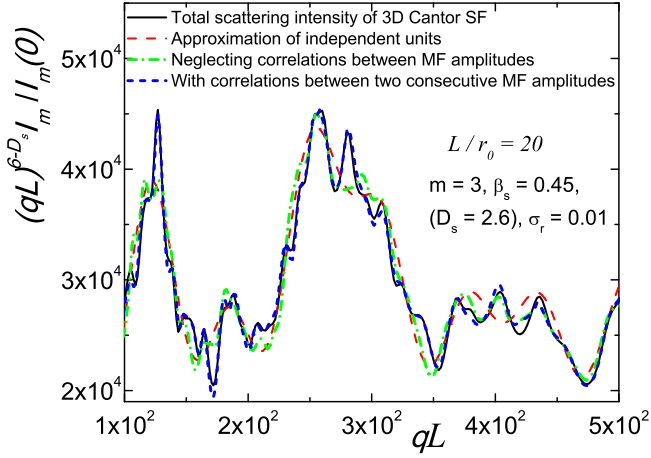}}
\caption{(Color online) The scaled scattering intensity $(qL)^{6-D_{\mathrm{s}}}I(q)$ of Fig.~\ref{fig:CSF_incoher_sum}a, but in a large scale to compare different approximations. The relative variance of polydispersity $\sigma_{\mathrm{r}}$ is equal to $0.01$. The approximation, taking into accounts the correlations between consecutive MF amplitudes (\ref{int_double_mass}) [dashed (blue) line], perfectly reproduces
the total scattering intensity [solid (black) line].
}
\label{fig:CompApp}
\end{figure}

It is clear that the more correlators in the total amplitude are taken into accounts, the better the approximation works. And conversely, the more correlators are neglected, the more interference minima and maxima disappear from the scattering intensity. The figure \ref{fig:CompApp} shows how the different approximations, discussed in Secs.~\ref{sec:SAS_SF_inMF}, \ref{sec:SAS_SF_in_units}, and \ref{sec:SAS_SF_in_pairs}, work. The most precise is Eq.~(\ref{int_double_mass}), perfectly reproducing the interference minima and maxima. We emphasize, however, that such an accuracy for the scattering intensity is not needed, because it is not observable in possible SAS experiments. For the given ratio $d/l=L/r_0=20\gg1$, even the approximation of spatially uncorrelated units works fairly well, reproducing fairly well the ``fine" structure of the SAS curve.

\section{\label{sec:rfs} Polydisperse components within random and deterministic surface fractals}

A deterministic surface fractal can be seen as a system of balls whose radii follow a discrete power-law distribution. Moreover, as discussed in Sec.~\ref{analysis}, positions of the balls are not important \emph{in the fractal region} for the rough structure of the scattering curve. Then one can expect that the only quantity, which is significant for the behavior of scattering intensity in the fractal region, is the exponent of power-low distribution. To show this, let us compare the discrete power-law distribution with continuous one having the same exponent.

It is important to make here a clear distinction between two types of polydispersities (log-normal vs. power-law) used in this paper: the log-normal polydispersity are related to the \textit{overall} sizes of \emph{different} Cantor surface fractals, which are assumed to be taken at random, while the power-law polydispersity is used here for describing the distribution of the ball radii \textit{inside one} surface fractal.

\begin{figure}[tb]
\centerline{\includegraphics[width=.6\columnwidth]{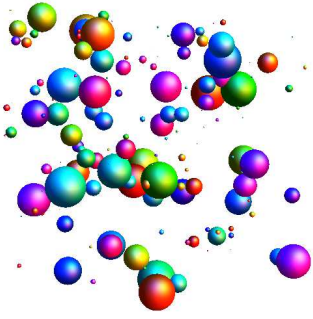}}
\caption{(Color online) A distribution of balls whose radii follow a power-law continuous distribution.}
\label{fig:balls}
\end{figure}

We consider further a system of non-overlapping balls in three-dimensional space (see Fig.~\ref{fig:balls}) with continuously distributed radii $r$, satisfying the condition $a \leqslant r \leqslant R$, where $a$ and $R$ are the smallest and largest radius of the balls, respectively. The number of balls $\d N(r)$ whose radii falls within the range ($r,r + \d r$)  is proportional to $\d r/r^{\tau}$ with $3<\tau<4$. An analog of finite iteration is the cutoff length $a$, for which only the balls of radii larger than $a$ are considered.

The exponent $\tau$ can easily be related to the fractal dimension of the \emph{combined surface area} of the balls (see Ref.~\cite{pfeifer83} and Appendix A in Ref.~\cite{chernyPRE11}). Let us prove that the total area of the sphere surfaces has the fractal dimension $D_{\mathrm{s}}=\tau-1$. According to the definition of Hausdorff (fractal) dimension, we should estimate the minimal number of \emph{balls} of radius $a$ needed to cover the set of \emph{spheres} when $a\to 0$. The minimal number of balls of radius $a$ needed to cover a \emph{sphere} of radius $r$ is proportional to $r^2/a^2$. Then the minimal number of balls for covering the system with a finite cutoff length $a$ is is given by the integral
\begin{equation}
N(a)\propto \frac{1}{a^2}\int_{a}^{R}\d r\, r^{2-\tau} \propto \frac{1}{a^{\tau-1}},
\label{eq:numberofballstocover}
\end{equation}
when $a\to 0$. Comparing this equation with the definition of Hausdorff dimension $N(a)\propto a^{-\ds}$
yields $\ds=\tau-1$ with $2 < D_{\mathrm{s}} < 3$.

Let us show how the above method of obtaining Hausdorff dimension works in some specific cases. If we consider the \emph{total volume} of the balls, their Hausdorff dimension is obviously equal to $D=3$. Indeed, the minimal number of balls of radius $a$ needed to cover a
\emph{ball} of radius $r$ is proportional to $r^3/a^3$, and we obtain in the same manner
\begin{equation}\label{nepball}
N(a)\propto \frac{1}{a^3}\int_{a}^{R}\d r\, r^{3-\tau} \propto a^{-3},
\end{equation}
because this integral converges at the lower limit of integration when $a\to0$ for $\tau<4$. Note that if $\tau>4$, the total volume of the balls diverges when $a\to0$, which means that such system of balls cannot be realized without overlaps between them at sufficiently low $a$. If $\tau<3$, the integral in Eq.~(\ref{eq:numberofballstocover}) converges for $a\to0$, which implies $N(a)\propto a^{-2}$, and $\ds=2$. This is in complete analogy with the Cantor surface fractal, whose surface dimension cannot be lower than $2$ (see the last paragraph of Sec.~\ref{subsec:Construction}).

Note that the positions of the balls in real space are supposed to be \emph{spatially uncorrelated}. In spite of this, the spatial correlations are still present in the system, because they are present in each ball composing this fractal. The power-law distribution of radii makes the resulting correlations to be of the fractal type.


In order to compare the discrete and continuous distributions, it is convenient to involve number of balls \emph{within a certain range}, which is not supposed to be small. It follows from the construction of the $m$th iteration of deterministic surface Cantor fractal (see Sec.~\ref{subsec:Construction}) that number of balls with radii $r'$ lying within $r' \leqslant r$ is given by the equation
\begin{equation}
N^{\mathrm{discr}}(r' \leqslant r)=\sum_{n=0}^{m}k^{n}\Theta(r-\beta_{s}^{n}r_{0}),
\label{eq:discretecdf}
\end{equation}
where $k=8$ and $\Theta(x)$ is the Heaviside step function, that is, $\Theta(x)=1$ for $x\geqslant 0$ and $\Theta(x)=0$ otherwise.

For the continuous distribution considered above [$\d N(r) \propto r^{-\tau}\d r$], we put first, the exponent $\tau=\ds+1$ with the fractal dimension $\ds$ being equal to the surface dimension of the Cantor surface fractal, $\ds=-\ln k/\ln\beta_{\mathrm{s}}$, and second, the total number of balls being equal to that of the $m$th Cantor fractal iteration $N_m$ [see Eq.~(\ref{Nm})]. We obviously have $N^{\mathrm{cont}}(r' \leqslant r)=C_1 r^{-\ds}+C_2$, where the unknown constants $C_1$ and $C_2$ can be found from the conditions $N^{\mathrm{cont}}(r' \leqslant a)=0$ and $N^{\mathrm{cont}}(r' \leqslant R)=N_m$. We derive
\begin{equation}
N^{\mathrm{cont}}(r' \leqslant r)= \frac{k^{m+1}-1}{k-1} \frac{(R/a)^{\ds}-(R/r)^{\ds}}{(R/a)^{\ds}-1}.
\label{eq:continuouscdf}
\end{equation}
This equation is valid for arbitrary $r$ lying between $a$ and $R$, otherwise $N^{\mathrm{cont}}$ is zero for $r\leqslant a$ and equal to $N_m=\frac{k^{m+1}-1}{k-1}$ when $r\geqslant R$. The parameters $R$ and $a$ should be chosen to ensure that the continuous distribution (\ref{eq:continuouscdf}) coincides with the discrete one (\ref{eq:discretecdf}) at the points $r_n=\beta_{\mathrm{s}}^{n}r_{0}$ for $n=0,\ldots,m$. Then they are given by
\begin{equation}
R=r_0,\quad a=\beta_{s}^{m+1}r_{0}.
\label{eq:Ra}
\end{equation}
Substituting the parameters (\ref{eq:Ra}) into Eq.~(\ref{eq:continuouscdf}) and using the relation $k\beta_{\mathrm{s}}^{\ds}=1$ finally yield
\begin{equation}
N^{\mathrm{cont}}(r' \leqslant r)= \frac{k^{m+1}-(r_0/r)^{\ds}}{k-1}.
\label{eq:continuouscdf1}
\end{equation}

\begin{figure}[tb]
\centerline{\includegraphics[width=0.9\columnwidth]{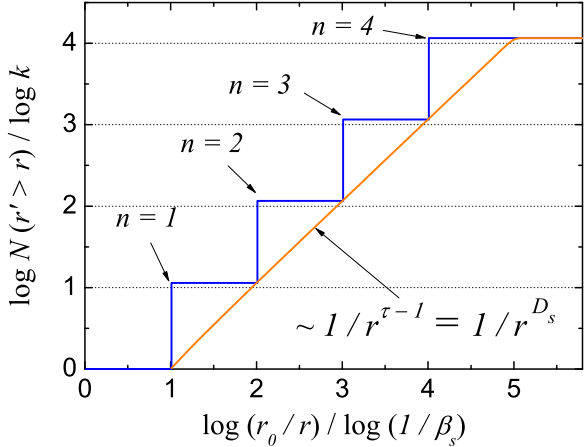}}
\caption{(Color online)
Distribution of balls composing the random (red) and deterministic (black) fractal of the fourth iteration on a double logarithm scale. Here $N(r'>r)$ is the number of balls with radii $r'$ obeying the condition $r'>r$. It is shown as a function of $1/r$. The step-like function indicates discreteness of the ball distribution \emph{within} the deterministic fractal.}
\label{fig:balls_distr}
\end{figure}

Because of the dominant contribution of small radii in the ``cumulative" distributions (\ref{eq:discretecdf}) and (\ref{eq:continuouscdf1}), it is more instructive to draw $N(r'>r)=N_m-N(r' \leqslant r)$ (that is, the number of balls with radii $r'$ obeying the condition $r'>r$) as a function of $1/r$. The double logarithm plot is shown in Fig.~\ref{fig:balls_distr}. One can see that the polydispersity distributions are alike in the power-law exponent and coincide at the ``corner" points.

Once $N(r' \leqslant r)$ is known explicitly as a function of $r$, the normalized distribution can be obtained by the relation $D_N(r)=(1/N_m) \d N/\d r$. We obtain from Eqs.~(\ref{eq:discretecdf}) and (\ref{eq:continuouscdf1}), respectively,
\begin{align}
D_{N}^{\mathrm{discr}}(r)&=\frac{k-1}{k^{m+1}-1} \sum_{j=0}^{m}k^{j}\delta(r-\beta_{s}^{j}r_{0}), \label{eq:discretepdf}\\
D_{N}^{\mathrm{cont}}(r)& =\frac{\ds}{k^{m+1}-1}\frac{r_0^{\ds}}{r^{\ds+1}}.  \label{eq:continuouspdf}
\end{align}
Here the well-known formula $\d\Theta(x)/\d x=\delta(x)$ is used. As expected, the discrete distribution function (\ref{eq:discretepdf}) is given by a sum of appropriately weighted Dirac's delta-functions. Equation (\ref{eq:continuouspdf}) is applicable for $\beta_{\mathrm{s}}^{m+1}r_{0} \leqslant r \leqslant r_{0}$, otherwise $D_{N}^{\mathrm{cont}}(r)=0$.

As is shown in Sec.~\ref{analysis}, the scattering intensity of a surface fractal \emph{in the fractal region} is a result of \textit{incoherent} diffraction of the units composing the fractal, namely, balls for the Cantor surface fractal or the random fractal with the power-low distribution. This means that the resulting intensity is a sum of the intensities of all balls composing the fractal. For a continuous distribution, the sum should be replaced by the corresponding integral. By analogy with Eq.~(\ref{eq:polyint}), we derive
\begin{equation}
I_m(q)=n\left| \Delta \rho \right|^{2}\int_{0}^{\infty}\d r F_{0}^{2}(qr) V_{\mathrm{b}}^2(r)D_N(r),
\label{eq:int_incoh}
\end{equation}
where $F_0$ is the form factor (\ref{eq:ballformfactor}) of ball of unit radius, $V_{\mathrm{b}}(r)=4\pi r^3/3$ is the volume of ball, and $D_N(r)$ is the normalized distribution given by Eq.~(\ref{eq:discretepdf}) or (\ref{eq:continuouspdf}). Certainly, for the discrete distributions, Eq.~(\ref{eq:int_incoh}) coincides with Eq.~(\ref{eq:intensitysumsquaresv2}) considered above up to a constant factor.

\begin{figure}[tb]
\centerline{\includegraphics[width=0.9\columnwidth]{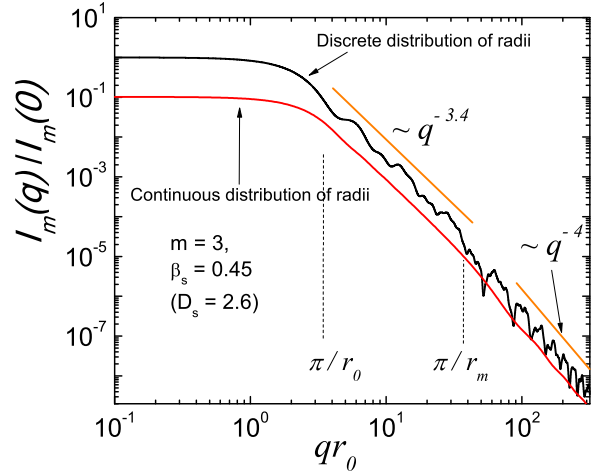}}
\caption{(Color online) The intensity of monodisperse scattering (\ref{eq:int_incoh}) from surface fractals with a discrete (black) [Eq.~(\ref{eq:discretepdf})] and continuous (red) [Eq.~(\ref{eq:continuouspdf})] power-law distribution of balls composing the fractals. The intensities are normalized to $I_m(0)$ of the discrete distribution, and the momentum transfer $q$ is represented in units of the largest ball radius $1/r_0$. The scattering from a surface fractal can be roughly explained in terms of  power-law distribution of sizes of objects composing the fractal. The distribution can be discrete (for deterministic fractals) or continuous (for random fractals).
}
\label{fig:SF_incoh_gen}
\end{figure}

The scattering intensities are shown in Fig.~\ref{fig:SF_incoh_gen}. As expected, the intensity curve is smoothed for the continuum power-low distribution (\ref{eq:continuouspdf}), but the scattering exponent $6-\ds$ is not changed, as well as the positions of the upper and lower edges of the fractal region.

It should be emphasized that the centers of the continuously distributed balls are assumed to be uncorrelated. The question arises whether the long-range correlations between the ball positions could contribute somehow into the fractal region or not. The deep analogy between the continuous and discreet power-law distributions can help us to answer the question. To this end, we consider the Cantor surface fractal with the same dimension $\ds=\tau-1$. So, the both systems (the Cantor surface fractal and the continuously distributed balls) have the same fractal dimension. Hence, according to the paper by Bale and Schmidt [5], the both fractals have to have the fractal region with the exponent $6-\ds$. The figure \ref{fig:SCF_int}a gives us the full range of correlations for the Cantor surface fractal including long- and short-ranged correlations. Only the range in green has the proper slope with the factor of $6-\ds$, and it is the fractal range that corresponds to the fractal range in Fig.~\ref{fig:SF_incoh_gen}. Black and red ranges (describing the long-ranged correlations, because small momenta are related to big distances in real space) do not show anything that vaguely resembles a fractal region, and this means that the long-range correlations hardly play a role in explaining the exponent $6-\ds$ (see also the arguments in Sec.~\ref{sec:SAS_MSF}).

\section{\label{sec:Conclusions}Conclusions}
We construct a deterministic surface fractal as a sum of three-dimensional mass Cantor sets at various iterations. We study its structural properties in momentum space and derive analytical expressions for monodisperse and polydisperse form factor, radius of gyration, and edges of the fractal regions.


We conclude that in general (with minor reservations discussed in the Introduction), \emph{any surface fractal can be represented as a sum of non-overlapping mass fractals}. This implies that the scattering amplitude of surface fractal can be written down as a sum of the amplitudes of composing mass fractals, see Eq.~(\ref{SF_ampl_MF}). This representation enables us to construct various approximations taking into account different correlations between the scattering amplitude of the objects that compose the surface fractal.

The roughest approximation is to consider the amplitudes of the composing primary objects being incoherent, which assumes that the spatial correlations between the primary objects are not important. This approximation always reproduces correctly the \emph{borders} of the fractal region for a surface fractal (see Sec.~\ref{sec:SAS_SF_generic}) and the rough structure of the scattering intensity. However, its fine structure, including tiny minima and maxima, is described well by this approximation only when $d/l\gg1$ (that is, \emph{the distance between objects is much larger than their size for each mass fractal composing the surface fractal}), otherwise more precise approximations are needed. One of them is Eq.~(\ref{int_incoh_MA}), which takes into consideration the correlations of the objects within each mass fractal, or Eq.~(\ref{int_double_mass}), which includes the correlations between \emph{pairs} of consecutive amplitudes of composing mass fractals. In this manner, one can always specify the fine structure of the scattering intensity of the surface fractal.

It is shown that when the spatial correlations between the primary objects are not important, small-angle scattering from a surface fractal can be described in the roughest approximation in terms of power-law distribution of sizes of objects composing the fractal (internal polydispersity). As is shown, the distribution of sizes $r$ of composing units obeys the power-law  $dN(r) \propto r^{-\tau}dr$, with $D_{\mathrm{s}}=\tau-1$; it is continuous for random surface fractals and discrete for deterministic surface fractals. Thus, the SAS from surface fractals can be roughly understood in terms of power-law type polydispersity. This could explain the physical nature of the exponent $\ds - 6$, found in Ref.~\cite{bale84} and solve the longstanding question whether the small-angle scattering from surface fractals can be explained in terms of polydispersity. The answer is ``yes", provided the polydispersity is of power-law type and the fine structure of the scattering intensity of the surface fractal is neglected.

The present analysis could also be helpful for extracting additional information from SAS data, such us the edges of the fractal region, the fractal iteration number and the scaling factor.

Modern SAS devices are able to measure the range of intensities within 5 or 6 orders of magnitude, while the measurable range of scattering vectors $q$ is limited to 3 orders. These limitations do not allow us to observe with a single experimental device all the properties obtained theoretically in this paper. In particular, one can measure only the initial part of the scattering intensity shown in Fig.~\ref{fig:SCF_int}a and miss the fractal and Porod regions. One can hope that rapid progress in experimental technics (see, e.g., Ref.~\cite{loh12}) will enhance our ability to observe the structure of matter at different scales.

\acknowledgments
The authors are grateful to Sergej Flach for useful remarks. The authors acknowledge financial support from JINR--IFIN-HH projects. A.I.K. acknowledges Russian program ``5Top100" of the Ministry of Education and Science of the Russian Federation.

\bibliography{sas_sf2}

\begin{thebibliography}{31}
\bibitem [{tbp()}]{tbp}%
  \BibitemOpen
  \href@noop {} {}\bibinfo {note} {To be published.}\BibitemShut {Stop}%
\end{thebibliography}

\end{document}